# Magneto-optical metamaterial

Mehdi Sadatgol[1], Mahfuzur Rahman[1], Ebrahim Forati[2], Miguel Levy[3], and Durdu Ö. Güney[1,*]

[1]Department of Electrical and Computer Engineering, Michigan Technological University, Houghton, MI 49931
[2] Department of Electrical and Computer Engineering, University of California San Diego, San Diego, CA 92093
[3] Department of Physics, Michigan Technological University, Houghton, MI 49931
*Corresponding author: dguney@mtu.edu

We propose a new class of metamaterials called magneto-optical metamaterials that offer enhanced angle of rotation in polarization compared to bulk magneto-optical materials. In the proposed approach, the permittivity tensor of a magneto-optical material is tailored by embedded wire meshes behaving as artificial plasma. We have shown that the angle of rotation in the magneto-optical metamaterial can be enhanced up to 9 times compared to bulk magneto-optical material alone while the polarization extinction ratio remains below -20dB and insertion loss is less than 1.5dB. © 2015 Optical Society of America

Non-reciprocal devices, such as isolators and circulators, are key components in today's laser based networks and optical links [1]. The magneto-optical Faraday Effect is at the heart of most non-reciprocal optical components. Zeeman or spin-orbit splitting of electronic energy levels in magnetic media results in dispersion differences between circularly-polarized optical modes of opposite helicity propagating in the magnetization direction [2]. This produces a nonreciprocal polarization rotation in linearly polarized light launched into the material (i.e. Faraday rotation). However, in the optical regime, the Faraday Effect is known to be rather weak. The specific Faraday rotation in bismuth- and cerium-substituted iron garnets, nonreciprocal materials of choice, ranges from 0.02°μm$^{-1}$ to ~1°μm$^{-1}$ at telecom wavelengths depending on substitution level [3,4]. Therefore, in order to achieve a desirable amount of rotation in the polarization, at least several tens of micrometers of MO material are required. Such a thick layer of MO material dramatically deteriorates the transmittance of the isolator or circulator [5]. Additionally, since typical fabrication methods such as microwave sputtering [6] or pulsed laser deposition [7] generally produce films up to few micrometers-thick depending on deposition time, several tens of micrometer-thick layers are also not favorable from a fabrication perspective. Liquid-phase epitaxy can achieve thicker films but bismuth substitution levels are rather limited, reducing achievable specific Faraday rotations and requiring even thicker films [8].

In order to achieve a large angle of rotation in a small length scale, resonant structures have been exploited such as a Fabry-Perot cavities filled with an MO material [9,10]. Since left-handed circularly polarized light (LCL) and right-handed circularly polarized light (RCL) propagate in MO materials with two different indices of refraction, the cavity resonates at different frequencies for the LCL and RCL. Consequently, around any of the resonance frequencies, the LCL and RCL components of a linearly polarized light beam acquire different phases as they pass through the cavity. A change in the relative phase difference of the LCL and RCL components of the linearly polarized light leads to a rotation in the angle of polarization. In order to obtain a polarization rotation by the angle of $\theta$, a phase difference of $\Delta\theta$ is required between the RCL and LCL components. However, this technique suffers from narrow band widths and small transmittance for structures with a single resonant cavity, making it unsuitable for optical communication applications which are essentially wideband [10]. Another problem is the introduction of ellipticity in the polarization, except at specific wavelengths [10]. These problems have been addressed theoretically through the introduction of multiple resonant cavities, but at the expense of complicating the fabrication significantly [11].

In this letter, we present a new method to enhance the angle of rotation of linearly polarized light in MO media. This approach, which we call magneto-optical metamaterial (MOM), benefits from the capability of the metamaterial to tailor the optical properties of the host MO media. In addition to the improvement in the angle of rotation, the MOM also allows for tuning the impedance and refractive index of the designed material. This would be useful, for instance, in designing low reflection polarization rotators. As another example, by adjusting the refractive index, one can make a single mode waveguide with glass or polymer core and MOM as the cladding, which otherwise is not feasible because refractive index of MO materials is typically higher than most glasses and polymers.

Optical properties of a typical MO material magnetized in the Z direction, is expressed with a scalar permeability, while the permittivity tensor is given as

$$\bar{\varepsilon}_{MO} = \begin{bmatrix} \varepsilon_\perp & ig & 0 \\ -ig & \varepsilon_\perp & 0 \\ 0 & 0 & \varepsilon_\parallel \end{bmatrix}. \quad (1)$$

$\varepsilon_\perp$ and $g$ describe the response of the material to electric field in the XY plane and $\varepsilon_\parallel$ is the permittivity in the direction of magnetization, the Z direction. Since the permeability is a scalar, the eigenvectors for the electromagnetic waves are the same as the eigenvectors of the permittivity tensor. With a straightforward algebraic calculation, one can find the MO material described with $\bar{\varepsilon}_{MO}$ supports LCL and RCL modes propagating in the Z direction with respective refractive indices of $n_+$ and $n_-$ given by

$$\begin{aligned} n_+ &= \sqrt{\mu(\varepsilon_\perp + g)}, \\ n_- &= \sqrt{\mu(\varepsilon_\perp - g)}. \end{aligned} \quad (2)$$

Consider a linearly polarized (LP) plane wave with angular frequency $\omega$ and time harmonic $e^{-j\omega t}$ is sent into a slab of MO material. After a distance $L$, the phase difference between the RCL and LCL components becomes $(n_+ - n_-)\frac{\omega}{c_0}L$, where $c_0$ is the speed of light in vacuum. If the amplitude of the output RCL and LCL components stay the same, the polarization of the total wave would still be linear and make an angle of $\Delta\theta$ with the direction of the polarization of the applied wave. The angle of rotation, $\Delta\theta$, is proportional to the difference between $n_+$ and $n_-$ as

$$\Delta\theta = \frac{1}{2}(n_+ - n_-)\frac{\omega}{c_0}L. \qquad (3)$$

Substituting Eq. (2) in Eq. (3) reveals through the plot in Fig. 1 that the angle of rotation of polarization is a monotonically decreasing function of $\varepsilon_\perp$. Therefore, one can improve the angle of rotation of polarization by decreasing $\varepsilon_\perp$. However, for a natural material it is not generally possible to tune the optical properties such as permittivity. On the other hand, metamaterials have provided unprecedented control of light since early 2000s giving birth to fascinating applications in imaging [12-20], solar photovoltaics [21], wireless communications [22,23], novel optical materials [24-27], and many more.

As we show below, metamaterials can also offer the degree of freedom to tune the permeability and permittivity of the magneto-optical material. In order to reduce the diagonal elements of the permittivity tensor, one can introduce an artificial plasma structure in the MO host medium. If such artificial plasma does not affect the off-diagonal elements of the resultant permittivity tensor but only influences the diagonal elements such as in isotropic medium [28], then one would expect that the artificial plasma embedded in the MO host medium would significantly enhance the angle of rotation of polarization above but near the plasma frequency. In the rest of the letter, we will verify this scenario.

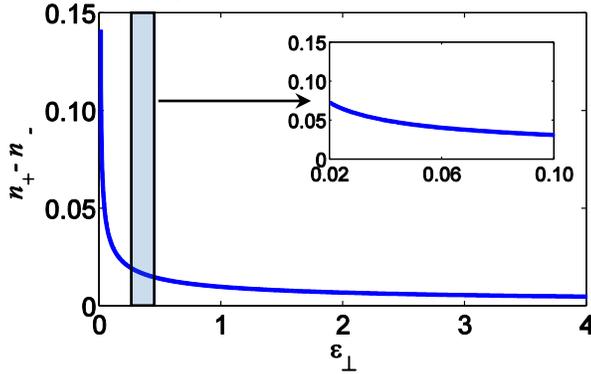

**Fig. 1.** $n_+ - n_-$ versus $\varepsilon_\perp$ for $g = 0.01$ and $\varepsilon_\perp$ from $g$ to 4. The subpanel shows a closer look for $\varepsilon_\perp$ from 0.02 to 0.10.

A potential implementation of such an artificial plasma embedded in an MO host medium is illustrated in Fig. 2. The artificial plasma is formed by conductive wires extending in the X and Y directions. The radius of the wires is $r$ and the separation between the adjacent parallel wires is equal to the size of the unit cell, $a$, for all the directions. We write the effective permittivity tensor for such a wire medium as [28]

$$\overline{\varepsilon}_{MOM} = 1 + \overline{\chi}^{POL+CON} \qquad (4)$$

where $\overline{\chi}^{POL+CON}$ is the effective susceptibility due to the polarization of bounded charges in the background MO medium and displacement of free electrons in the conducting wires. $\overline{\varepsilon}_{MOM}$ is the effective permittivity tensor for the MOM. Following the detailed procedure for the homogenization of the wire medium in Ref. 29, we can find the effective permittivity tensor in terms of topological parameters and material properties as

$$\overline{\varepsilon}_{MOM} \approx \overline{\varepsilon}_{MO} - \kappa \begin{bmatrix} 1 & 0 & 0 \\ 0 & 1 & 0 \\ 0 & 0 & 0 \end{bmatrix}; \qquad (5)$$

$$\kappa = \left(\frac{k_0^2}{k_p^2} - \frac{1}{f_v}\frac{1}{\varepsilon_m - \varepsilon_\perp}\right).$$

In Eq. (5), $\overline{\varepsilon}_{MO}$ is the permittivity tensor of the host MO medium [see Eq. (1)], $k_0$ is the wavenumber in vacuum, $f_v = \frac{\pi r^2}{a^2}$ is volumetric filling factor for the wires, $k_p$ is the plasma wavenumber $\left((k_p a)^2 \approx 2\pi/\ln(a^2/4r(a-r))\right)$, and $\varepsilon_m$ is the relative permittivity of the metallic wires. Note that a few assumptions were made in deriving Eq. (5). As such, $f_v$ is very small, the spatial dispersion is neglected.

As it is clear from Eq. (5), one can tune the diagonal elements of the MOM permittivity tensor by selecting an appropriate value for $\kappa$, which is a function of optical properties of the constituent materials, radius of the conducting wires, and size of the unit cell. The largest improvement in the angle of rotation of polarization occurs at the frequency, where $\varepsilon_\perp - \kappa = g$, because this is the smallest possible diagonal element for the MOM. When $\varepsilon_\perp - \kappa < g$, which happens at the lower frequencies, the eigenmodes of the MOM start to become evanescent (i.e., this can be easily seen from Eq. (2) by replacing $\varepsilon_\perp$ with $\varepsilon_\perp - \kappa$, since Eq. (1) and Eq. (5) have similar tensor forms). However, if impedance matching is of concern, $\kappa$ needs to be selected such that the impedance of the MOM, $\sqrt{1/(\varepsilon_\perp - \kappa)}$, approaches the impedance of the surrounding dielectric medium at the desired frequency range.

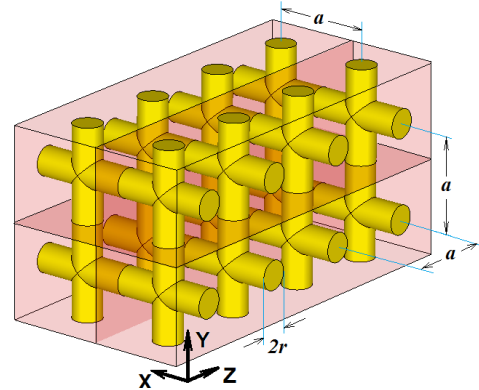

**Fig. 2.** MOM consists of conductive wires (yellow) in MO background (red). The radius of the wire is $r$ and the separation between the adjacent wires is $a$.

Here, we focus on the enhancement in the angle of rotation at telecom wavelength ($\lambda_0 = 1.55\mu m$). The design parameters are the radius of the wires, the length of the unit cell, and the constituent material properties. The length of the unit cell needs to be much smaller than the working wavelength in order to keep the effective medium approximation valid. We used gold for the wires, which are embedded inside bismuth-substituted yttrium iron garnet (Bi:YIG) with $\varepsilon_\perp = 5.7$, $g = 0.01$ and $\mu = 1$ [30]. Bi:YIG is selected for the host MO medium because of its high specific Faraday rotation at telecom frequency. Once we have determined the length of the unit cell and the constituent materials, Eq. (5) gives the appropriate radius of the wires for the maximum angle of rotation at $\lambda_0 = 1.55\mu m$. Selecting the length of unit cell to be $130\text{nm}$, we find the required radius of the wires as $r = 17\text{nm}$.

The MOM shown in Fig. 2 with the designed parameters were simulated in CST Microwave Studio [31]. The structure extends infinitely in the X and Y directions and has finite length along the Z direction which is the direction of propagation. The gold wires were described by the Drude model with plasma frequency $f_p = 2150\text{THz}$ and collision frequency $f_c = 6.5\text{THz}$ [32]. The length of the structure in the direction of propagation, was swept from $4a$ to $20a$ with $2a$ steps. The maximum angle of rotation for every step is recorded and displayed in Fig. 3. As can be seen in the figure the simulation data points follow approximately a straight line which implies that the MOM is almost linearly scalable in the direction of propagation. The amount of rotation in polarization as a linearly polarized wave propagates in bulk B-YIG, given by Eq. (3), is also shown in Fig. 3. In order to compare the MOM and bulk MO materials, we have fitted a line to the data points by minimizing the square of the errors between the fitted line and the data points. There are two different ways to find the enhancement in the rotation of polarization. The first one is to directly divide the data points by the corresponding values of bulk B-YIG, and the next one is to compare the slope the MOM line with B-YIG line in Fig. 3. Although for short structures the two methods give different values for enhancement in the rotation of polarization, they agree as the length of the structures in the direction of propagation increases. This behavior is not surprising since metamaterials become length independent as the overall length of metamaterial increases [33-36]. The maximum enhancement in the rotation of polarization regardless of the ellipticity of the output light is around 6 when the metamaterial is as short as $1\mu m$ and it approaches asymptotically to about 12 when the length of the metamaterial is larger than $2\mu m$. We will defer the discussion of ellipticity and its influence on the enhancement until the end of the letter.

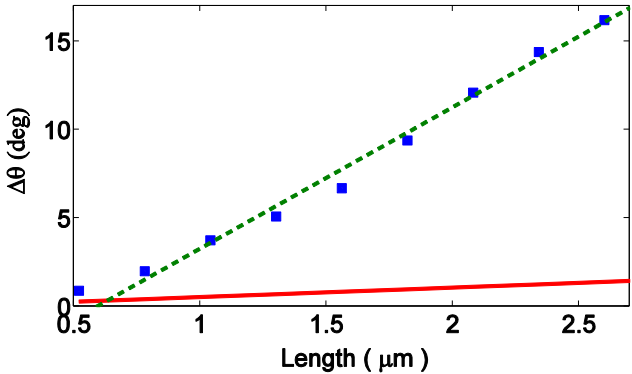

**Fig. 3.** The angle of rotation vs. length for the MOM and bulk B-YIG. Filled squares show the data points for the MOM obtained from the simulations. Dashed line indicates the fitted line to the data points. The solid line shows the angle of rotation for the bulk B-YIG.

In the following, we will discuss retrieved optical parameters of the MOM. There are motivations to retrieve the permittivity of the proposed metamaterial from the results of the numerical simulation. First, to independently examine the underlying idea of MOM discussed earlier, second is to verify the analytical expression for permittivity given in Eq. (5). To the best of our knowledge, there is no prior work presenting retrieval procedure for MO materials. In order to retrieve optical properties of the MOM, we have used an approach similar to the conventional retrieval procedure based on transfer matrix approach [37]. This enables us to find the refractive indices for the LCL and RCL, $n_+$ and $n_-$, respectively. Once $n_+$ and $n_-$ are obtained, the elements $\varepsilon_{\perp,eff}$ and $g_{eff}$ of the effective permittivity tensor can be calculated from Eq. (2). However, before we apply this method to the MOM, we need to first ensure that LCL and RCL are the eigenmodes of the material. The effective permittivity tensor of the MOM in Eq. (5) has similar tensor form as the permittivity tensor in Eq. (1), thus one should expect that the eigenmodes for the MOM should be also LCL and RCL. To verify this, we sent LCL and RCL waves separately to the MOM and found that the reflected and transmitted waves in both cases have the same form as the input wave.

The resultant effective parameters, $\varepsilon_{\perp,eff}$ and $g_{eff}$, corresponding to the MOM are shown in Fig. 4. One can see in the figure that the plasma frequency of the MOM, the frequency which $\varepsilon_\perp = \kappa$, is about 188THz. Note that the retrieved $g_{eff}$, is about 10% less than what is predicted by Eq. (5). In order to reconcile this discrepancy it is important to recall that the elements of the permittivity tensor are volumetric averages. In the derivation of Eq. (5) the volume of conducting wires has been neglected. Since conducting wires do not contribute to the gyration, Eq. (5) overestimates $g_{eff}$, by about a factor of $2f_v$ (i.e., $2f_v$ is the ratio of the volume of conducting wires to the total volume of the metamaterial). In the metamaterial under study, about 10% of the total volume is gold. Therefore, the actual value of the $g_{eff}$, is about 10% smaller than what Eq. (5) predicts.

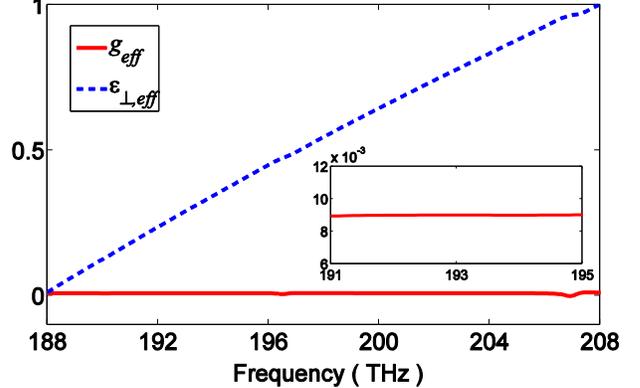

**Fig. 4.** Retrieved effective permittivity tensor elements for the MOM described in Fig. 2 with 16 unit cells in the direction of propagation. (Blue dashed line) $\varepsilon_{\perp,eff}$, and (red) $g_{eff}$. The subpanel shows a closer look of $g_{eff}$ for frequency from 191 THz to 196THz.

Next, we apply a linearly polarized light to the MOM and study how the angle of rotation of polarization changes with the frequency. Fig. 5 shows the angle of rotation of polarization (i.e., normalized with respect to the maximum angle of rotation) versus frequency. As we expected, the amount of rotation in the polarization is maximum after the plasma frequency and decreases as the frequency increases.

MO medium, in general, and the proposed MOM, in particular, do not respond to RCL and LCL in the same way. For structures which contain MO materials, it is common to observe different transmittance for these two states of polarization. Consequently, a linearly polarized light, which is a combination of LCL and RCL with equal amplitudes, does not maintain linearly polarization as it propagates through an MO medium. More precisely, polarization slightly deviates from linear to elliptical polarization. Polarization extinction ratio (PER), which is defined as the ratio of the minor axis of the polarization ellipse to the major axis, quantifies the ellipticity of the light. PER has a major impact on the level of isolation between ports in an isolator or a circulator. PER along with transmittance define the working bandwidth of MOM for a particular application. Fig. 5 shows PER in dB, transmittance, and angle of rotation versus frequency. For instance, consider the isolator at the telecom frequency in Ref. 38, where the PER is required to be less than 20dB. Then, the working bandwidth of our MOM is more than 2THz, which covers more than a half of the C-band (1530-1565nm ). The shaded region in Fig. 5 shows the operating bandwidth that corresponds to PER<-20dB. The transmittance inside the working window is about 0.7 (1.5dB insertion loss) and the enhancement in the angle of rotation varies from a factor of 9 to 4.5 compared to bulk MO material.

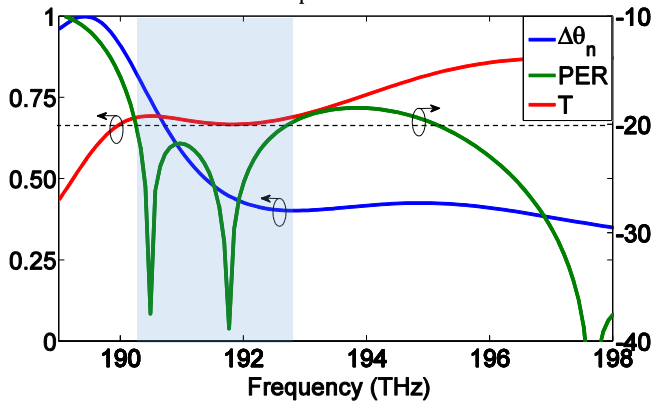

**Fig. 5.** (blue) normalized angle of rotation ($\Delta\theta_n$), (green) PER in dB, and (red) Transmittance (T) for the MOM described in Fig. 2 with 16 unit cells in the direction of propagation. The shaded region demonstrates the working bandwidth corresponding to PER<-20dB.

Although the artificial plasma considered in this letter is the well-known wire medium, any subwavelength design of artificial plasma can be used to obtain similar results. The MOM structure could be modified to accommodate various design and fabrication requirements. As an example, one can consider a structure consisting of a periodic stack of thin metal film and B-YIG layer. The multilayer structure could be easily fabricated by traditional fabrication methods such as RF sputtering and magnetron sputtering. We have numerically studied the multilayer structure and obtained similar permittivity tensor elements as in Fig. 4 for the same gold to B-YIG volumetric ratio as the designed wire medium based MOM.

In summary, we have proposed a new class of metamaterials capable of mimicking the optical properties of natural MO materials with the added advantages of enhancing the angle of rotation of polarization and minimizing the reflection from interfaces with lower index dielectrics. We verified the analytical optical properties of the proposed MOM through simulations and a numerical procedures devised specifically for retrieving the effective permittivity tensor for MOMs.

**Funding**. National Science Foundation (NSF) (ECCS-1202443).

## References

1. D. Jalas, A. Petrov, M. Eich, W. Freude, S. Fan, Z. Yu, R. Baets, M. Popovic, A. Melloni, J. D. Joannopoulos, M. Vanwolleghem, C. R. Doerr, and H. Renner, Nat. Photonics **7**, 579 (2013).
2. P. N. Argyres, Phys. Rev. **97**, 334 (1955).
3. M. Chandra Sekhar, M. R. Singh, S. Basu, Sai Pinnepalli, Opt. Express **20**, 9624 (2012).
4. H. Yokoi, Y. Shoji, and T. Mizumoto, Jpn. J. Appl. Phys. **43**, 5871 (2004).
5. J-L. Rehspringer, J. Bursik, D. Niznansky, and A. Klarikova, J. Magnetism and Magnetic Materials **211** (1), 291 (2000).
6. T. Boudiar, B. Payet-Gervy, M-F. Blanc-Mignon, J-J. Rousseau, M. L. Berre, and H. Joisten, J. Magnetism and Magnetic Materials **284**, 77 (2004).
7. S. Kahl, and A. M. Grishin, J. Appl. Phys. **93** (10), 6945 (2003).
8. "Epitaxial Garnet Films for Non-Reciprocal Magnetooptic Devices," V. J. Fratello and R. Wolfe, Chapter 3 in Magnetic Film Devices, edited by M. H. Francombe and J. D. Adam, Volume 4 of Handbook of Thin Film Devices: Frontiers of Research, Technology and Applications (Academic Press, 2000).
9. M. Inoue, and T. Fujii, J. Appl. Phys. **81** (8), 5659 (1997).
10. M. J. Steel, M. Levy, and R. M. Osgood Jr., Photonics Tech. Lett., IEEE **12** (9), 1171 (2000).
11. M. Levy, H. C. Yang, M. J. Steel, and J. Fujita, J. Lightwave Technol. **19**, 1964 (2001).
12. V. G. Veselago, Sov. Usp., **10**, 509 (1968).
13. T. Xu, A. Agrawal, M. Abashin, K. J. Chau, and H. J. Lezec, Nature **497**, 470 (2013).
14. J. B. Pendry, Phys. Rev. Lett. **85**, 3966 (2000).
15. M. Sadatgol, S. K. Ozdemir, L. Yang, and D. O. Guney, Phys. Rev. Lett. **115**, 035502 (2015).
16. Z. Jacob, L. V. Alekseyev, and E. Narimanov, Opt. Express **14**, 8247 (2006).
17. Z. Liu, H. Lee, Y. Xiong, C. Sun, and X. Zhang, Science **315**, 1686 (2007).
18. X. Zhang and Z. Liu, Nat. Mater. **7**, 435 (2008).
19. J. Rho, Z. Ye, Y. Xiong, X. Yin, Z. Liu, H. Choi, G. Bartal, and X. Zhang, Nat. Commun. **1**, 143 (2010).
20. J. Sun, M. I. Shalaev, and N. M. Litchinitser, Nat. Commun. **6**, 7201 (2015).
21. A. Vora, J. Gwamuri, N. Pala, A. Kulkarni, J. M. Pearce, and D. O. Guney, Sci. Rep. **4**, 4901 (2014).
22. I. Bulu, H. Caglayan, K. Aydin, and E. Ozbay, New J. Phys. **7**, 223 (2005).
23. H. Odabasi, F. Teixeira, and D. O. Guney, J. Appl. Phys. **113**, 084903 (2013).
24. J. Valentine, S. Zhang, T. Zentgraf, E. Ulin-Avila, D. A. Genov, G. Bartal, and X. Zhang, Nature **455**, 376 (2008).
25. N. I. Landy, S. Sajuyigbe, J. J. Mock, D. R. Smith, and W. J. Padilla, Phys. Rev. Lett. **100**, 207402 (2008).
26. K. Aydin, V. E. Ferry, R. M. Briggs, and H. A. Atwater, Nature Commun. **2**, 517 (2011).
27. M. I. Aslam and D. O. Guney, Progress In Electromagnetics Research B **47**, 203 (2013).
28. G. W. Hanson, E. Forati, and M. G. Silveirinha, Antennas and Propagation, IEEE Transactions on **60** (9), 4219 (2012).
29. M. G. Silveirinha, Phys. Rev. B **79** (3), 035118 (2009).
30. M. Y. Chern, F. Y. Lo, D. R. Liu, K. Yang, and J. S. Liaw, Jpn. J. Appl. Phys. **38**(12R), 6687 (1999).
31. https://www.cst.com/Products/CSTMWS
32. M. A. Ordal, L. L. Long, R. J. Bell, S. E. Bell, R. R. Bell, R. W. Alexander, and C. A. Ward, Appl. Opt. **22**(7), 1099 (1983).
33. T. Koschny, P. Markoš, Eleftherios N. Economou, D. R. Smith, D. C. Vier, and C. M. Soukoulis, Phys. Rev. B **71**(24), 245105 (2005).
34. D. O. Guney, Th. Koschny, M. Kafesaki, C. M. Soukoulis, Opt. Lett. **34**, 506 (2009).
35. D. O. Guney, Th. Koschny, and C. M. Soukoulis, Opt. Express **18**, 12348 (2010).
36. M. I. Aslam and D. O. Guney, J. Opt. Soc. Am. B **29**, 2839 (2012).
37. D. R. Smith, D. C. Vier, Th Koschny, and C. M. Soukoulis, Phys. Rev. E **71** (3), 036617 (2005).
38. L. Bi, J. Hu, P. Jiang, D. H. Kim, G. F. Dionne, L. C. Kimerling, and C. A. Ross, Nat. Photonics **5**, 758 (2011).

# References

1. D. Jalas, A. Petrov, M. Eich, W. Freude, S. Fan, Z. Yu, R. Baets, M. Popovic, A. Melloni, J. D. Joannopoulos, M. Vanwolleghem, C. R. Doerr, and H. Renner, "What is – and what is not – an optical isolator," Nat. Photonics **7**, 579 (2013).
2. P. N. Argyres, "Theory of the Faraday and Kerr effects in ferromagnetics," Phys. Rev. **97**, 334 (1955).
3. M. Chandra Sekhar, M. R. Singh, S. Basu, Sai Pinnepalli, "Giant Faraday rotation in $Bi_xCe_{3-x}Fe_5O_{12}$ epitaxial garnet films," Opt. Express **20**, 9624 (2012).
4. H. Yokoi, Y. Shoji, and T. Mizumoto, "Calculation of nonreciprocal phase shift in magnetooptic waveguide with Si guiding layer," Jpn. J. Appl. Phys. **43**, 5871 (2004).
5. J-L. Rehspringer, J. Bursik, D. Niznansky, and A. Klarikova, "Characterisation of bismuth-doped yttrium iron garnet layers prepared by sol-gel process," J. Magnetism and Magnetic Materials **211** (1), 291 (2000).
6. T. Boudiar, B. Payet-Gervy, M-F. Blanc-Mignon, J-J. Rousseau, M. L. Berre, and H. Joisten, "Magneto-optical properties of yttrium iron garnet (YIG) thin films elaborated by radio frequency sputtering," J. Magnetism and Magnetic Materials **284**, 77 (2004).
7. S. Kahl, and A. M. Grishin, "Pulsed laser deposition of $Y_3Fe_5O_{12}$ films on garnet substrates," J. Appl. Phys. **93** (10), 6945 (2003).
8. "Epitaxial Garnet Films for Non-Reciprocal Magnetooptic Devices," V. J. Fratello and R. Wolfe, Chapter 3 in Magnetic Film Devices, edited by M. H. Francombe and J. D. Adam, Volume 4 of Handbook of Thin Film Devices: Frontiers of Research, Technology and Applications (Academic Press, 2000).
9. M. Inoue, and T. Fujii, "A theoretical analysis of magneto-optical Faraday effect of YIG films with random multilayer structures," J. Appl. Phys. **81** (8), 5659 (1997).
10. M. J. Steel, M. Levy, and R. M. Osgood Jr., "High transmission enhanced Faraday rotation in one-dimensional photonic crystals with defects," Photonics Tech. Lett., IEEE **12** (9), 1171 (2000).
11. M. Levy, H. C. Yang, M. J. Steel, and J. Fujita, "Flat-top response in one-dimensional magnetic photonic bandgap structures with Faraday rotation enhancement," J. Lightwave Technol. **19**, 1964 (2001).
12. V. G. Veselago, "The electrodynamics of substances with simultaneously negative values of permittivity and permeability," Sov. Usp., **10**, 509 (1968).
13. T. Xu, A. Agrawal, M. Abashin, K. J. Chau, and H. J. Lezec, "All-angle negative refraction and active flat lensing of ultraviolet light," Nature **497**, 470 (2013).
14. J. B. Pendry, "Negative refraction makes a perfect lens," Phys. Rev. Lett. **85**, 3966 (2000).
15. M. Sadatgol, S. K. Ozdemir, L. Yang, and D. O. Guney, "Plasmon injection to compensate and control losses in negative index metamaterials," Phys. Rev. Lett. **115**, 035502 (2015).
16. Z. Jacob, L. V. Alekseyev, and E. Narimanov, "Optical hyperlens: far-field imaging beyond the diffraction limit," Opt. Express **14**, 8247 (2006).
17. Z. Liu, H. Lee, Y. Xiong, C. Sun, and X. Zhang, "Far-field optical hyperlens magnifying sub-diffraction-limited objects," Science **315**, 1686 (2007).
18. X. Zhang and Z. Liu, "Superlenses to overcome the diffraction limit," Nat. Mater. **7**, 435 (2008).
19. J. Rho, Z. Ye, Y. Xiong, X. Yin, Z. Liu, H. Choi, G. Bartal, and X. Zhang, "Spherical hyperlens for two-dimensional sub-diffractional imaging at visible frequencies," Nat. Commun. **1**, 143 (2010).
20. J. Sun, M. I. Shalaev, and N. M. Litchinitser, "Experimental demonstration of a non-resonant hyperlens in the visible spectral range," Nat. Commun. **6**, 7201 (2015).
21. A. Vora, J. Gwamuri, N. Pala, A. Kulkarni, J. M. Pearce, and D. O. Guney, "Exchanging Ohmic losses in metamaterial absorbers with useful optical absorption for photovoltaics," Sci. Rep. **4**, 4901 (2014).
22. I. Bulu, H. Caglayan, K. Aydin, and E. Ozbay, "Compact size highly directive antennas based on the SRR metamaterial medium," New J. Phys. **7**, 223 (2005).
23. H. Odabasi, F. Teixeira, and D. O. Guney, "Electrically small, complementary electric-field-coupled resonator antennas," J. Appl. Phys. **113**, 084903 (2013).
24. J. Valentine, S. Zhang, T. Zentgraf, E. Ulin-Avila, D. A. Genov, G. Bartal, and X. Zhang, "Three-dimensional optical metamaterial with a negative refractive index," Nature **455**, 376 (2008).
25. N. I. Landy, S. Sajuyigbe, J. J. Mock, D. R. Smith, and W. J. Padilla, "Perfect metamaterial absorber," Phys. Rev. Lett. **100**, 207402 (2008).
26. K. Aydin, V. E. Ferry, R. M. Briggs, and H. A. Atwater, "Broadband polarization-independent resonant light absorption using ultrathin plasmonic super absorbers," Nature Commun. **2**, 517 (2011).
27. M. I. Aslam and D. O. Guney, "On negative index metamaterial spacers and their unusual optical properties," Progress In Electromagnetics Research B **47**, 203 (2013).
28. G. W. Hanson, E. Forati, and M. G. Silveirinha, "Modeling of spatially-dispersive wire media: transport representation, comparison with natural materials, and additional boundary conditions," Antennas and Propagation, IEEE Transactions on **60** (9), 4219 (2012).
29. M. G. Silveirinha, "Artificial plasma formed by connected metallic wires at infrared frequencies," Phys. Rev. B **79** (3), 035118 (2009).
30. M. Y. Chern, F. Y. Lo, D. R. Liu, K. Yang, and J. S. Liaw, "Red-shift of Faraday rotation in thin films of completely bismuth-substituted iron garnet $Bi_3Fe_5O_{12}$," Jpn. J. Appl. Phys. **38**(12R), 6687 (1999).
31. https://www.cst.com/Products/CSTMWS
32. M. A. Ordal, L. L. Long, R. J. Bell, S. E. Bell, R. R. Bell, R. W. Alexander, and C. A. Ward, "Optical properties of the metals Al, Co, Cu, Au, Fe, Pb, Ni, Pd, Pt, Ag, Ti, and W in the infrared and far infrared," Appl. Opt. **22**(7), 1099 (1983).
33. T. Koschny, P. Markoš, Eleftherios N. Economou, D. R. Smith, D. C. Vier, and C. M. Soukoulis, "Impact of inherent periodic structure on effective medium description of left-handed and related metamaterials," Phys. Rev. B **71**(24), 245105 (2005).
34. D. O. Guney, Th. Koschny, M. Kafesaki, C. M. Soukoulis, "Connected bulk negative index photonic metamaterials," Opt. Lett. **34**, 506 (2009).
35. D. O. Guney, Th. Koschny, and C. M. Soukoulis, "Intra-connected three-dimensionally isotropic bulk negative index photonic metamaterial," Opt. Express **18**, 12348 (2010).
36. M. I. Aslam and D. O. Guney, "Dual-band, double-negative, polarization-independent metamaterial for the visible spectrum," J. Opt. Soc. Am. B **29**, 2839 (2012).
37. D. R. Smith, D. C. Vier, Th Koschny, and C. M. Soukoulis, "Electromagnetic parameter retrieval from inhomogeneous metamaterials," Phys. Rev. E **71** (3), 036617 (2005).
38. L. Bi, J. Hu, P. Jiang, D. H. Kim, G. F. Dionne, L. C. Kimerling, and C. A. Ross, "On-chip optical isolation in monolithically integrated non-reciprocal optical resonators," Nat. Photonics **5**, 758 (2011).